\documentclass[aps,pra,showkeys,showpacs,superscriptaddress,reprint]{revtex4-1}
	
	\usepackage{setspace}
	\usepackage{amsmath}    
	\usepackage{graphicx}   
	\usepackage{verbatim}   
	\usepackage{color}      
	\usepackage{subfigure}  
	\usepackage{hyperref}   
	\usepackage{dcolumn}   
		
\makeindex

\begin{document}

\pagestyle{myheadings}

\title{Magic wavelengths for the $2 \ ^3S \to 2 \ ^1S$ transition in helium}

\author{R.P.M.J.W. \surname{Notermans}}
\author{R.J. \surname{Rengelink}}
\affiliation{LaserLaB, Department of Physics and Astronomy, VU University, De Boelelaan 1081, 1081 HV Amsterdam, Netherlands}
\author{K.A.H. \surname{van Leeuwen}}
\affiliation{Department of Applied Physics, Eindhoven University of Technology, PO Box 513, 5600 MB Eindhoven, Netherlands}
\author{W. \surname{Vassen}}
\email{w.vassen@vu.nl}
\affiliation{LaserLaB, Department of Physics and Astronomy, VU University, De Boelelaan 1081, 1081 HV Amsterdam, Netherlands}


\begin{abstract}
We have calculated ac polarizabilities of the $2 \ ^3S$ and $2 \ ^1S$ states of both $^4$He and $^3$He in the range 318~nm to 2.5~$\mu$m and determined the magic wavelengths at which these polarizabilities are equal for either isotope. The calculations, only based on available \textit{ab initio} tables of level energies and Einstein A coefficients, do not require advanced theoretical techniques. The polarizability contribution of the continuum is calculated using a simple extrapolation beyond the ionization limit, yet the results agree to better than $1\%$ with such advanced techniques. Several promising magic wavelengths are identified around 320~nm with sufficient accuracy to design an appropriate laser system. The extension of the calculations to $^3$He is complicated due to the additional hyperfine structure, but we show that the magic wavelength candidates around 320~nm are predominantly shifted by the isotope shift.
\end{abstract}


\maketitle

\section{INTRODUCTION}
In recent years a growing number of experimental tests of QED in atomic physics have surpassed the accuracy of theory, allowing new determinations of fundamental constants. High-precision spectroscopy in atomic hydrogen has been achieved with sufficient accuracy to allow a determination of the proton size from QED calculations \cite{Mohr1}, and spectroscopy in muonic hydrogen has allowed an even more accurate determination \cite{Pohl2,Antognini1}. Interestingly, the muonic hydrogen result currently differs by $7\sigma$ from the proton size determined by hydrogen spectroscopy and electron-proton collision experiments. So far there has not been a satisfying explanation for this discrepancy, which is aptly named the proton radius puzzle \cite{Pohl1}. Research in this field has expanded to measurements in muonic helium ions, a hydrogenic system which has a different nuclear charge radius \cite{Nebel1}. As this work is done for both naturally occurring isotopes of helium ($^4$He and $^3$He), the absolute charge radii of the $\alpha$-particle and the helion may be determined at an aimed relative precision of $3 \times 10^{-4}$ (0.5~attometer), providing a very interesting testing ground for both QED and few-body nuclear physics.

Parallel to these developments, high-precision spectroscopy in neutral helium has become an additional contribution to this field in recent years. Although QED calculations for three-body systems are not as accurate as for hydrogen(ic) systems, mass-independent uncertainties cancel when considering the isotope shift \cite{Drake4,Morton1}. Therefore isotope-shift measurements in neutral helium can provide a crucial comparison of the nuclear charge radius difference determined in the muonic helium ion and planned electronic helium ion measurements.

High-precision spectroscopy in helium is a well-established field, and transitions ranging from wavelengths of 51~nm to 2058~nm \cite{CancioPastor1,Zelevinski1,Eyler1,Borbely1,Kandula1,Smiciklas1,Rooij1,CancioPastor2,Luo1,*Luo1_erratum,Luo2,Notermans1} have been measured in recent years both from the ground state and from several (metastable) excited states. Only two transitions have been measured in both helium isotopes with sufficient precision for accurate nuclear charge radius difference determinations. The $2 \ ^3S \to 2 \ ^3P$ transition at 1083~nm \cite{CancioPastor2} and the doubly-forbidden $2 \ ^3S \to 2 \ ^1S$ transition at 1557~nm \cite{Leeuwen1,Rooij1} are measured at accuracies exceeding $10^{-11}$, providing an extracted nuclear charge radius difference with 0.3\% and 1.1\% precision, respectively. Interestingly, the determined nuclear charge radius differences from both experiments currently disagree by $4\sigma$ \cite{CancioPastor2}.

In order to determine the nuclear charge radius difference with a precision comparable to the muonic helium ion goal, we aim to measure the $2 \ ^3S \to 2 \ ^1S$ transition with sub-kHz precision. One major improvement to be implemented is the elimination of the ac Stark shift induced by the optical dipole trap (ODT) in which the transition is measured. Many high-precision measurements involving optical (lattice) traps solve this problem by implementation of a so-called magic wavelength trap \cite{Poli1,Ludlow1}. In a magic wavelength trap the wavelength is chosen such that the ac polarizabilities of both the initial and final states of the measured transition are equal, thereby cancelling the differential ac Stark shift. 

In this paper we calculate the wavelength-dependent (ac) polarizabilities of both metastable $2 \ ^3S$ (lifetime $\approx 7800$~s) and $2 \ ^1S$ (lifetime $\approx 20$~ms)  states and identify wavelengths at which both are equal for either $^4$He or $^3$He. Generally one will find multiple magic wavelengths over a broad wavelength range, but our goal is to identify the most useful magic wavelength for our experiment. Currently \cite{Rooij1,Notermans1} we employ a 1557~nm ODT at a power of a few 100~mW, providing a trap depth of a few $\mu$K and a scattering lifetime of $>$~100~s (the actual lifetime in the trap is limited to 10's of seconds due to background collisions). A good overview on calculating trap depths and scattering rates in ODTs is given in \cite{Grimm1}, and the specific calculations for our ODT are discussed in the Appendix. For our future magic wavelength trap we need to produce a similar trap depth with sufficient laser power at that wavelength. Furthermore, the scattering rate should be low enough to have a lifetime of at least a few seconds, providing enough time to excite the atoms with a 1557-nm laser.
 
The purpose of this paper is to show that it is possible to calculate magic wavelengths with sufficient accuracy to design an appropriate laser system solely based on \textit{ab initio} level energies and Einstein A coefficients without having to resort to advanced theoretical techniques \cite{Yan1,Mitroy1}. Based on the calculations reported here, we are currently building a laser system at 319.82~nm with a tuning range of 300~GHz based on similar designs \cite{Wilson1,Lo1}.

The polarizabilities for the $2 \ ^3S$ and $2 \ ^1S$ states of $^4$He are presented over a wavelength range from 318~nm to 2.5~$\mu$m. In this range all magic wavelengths including estimated required ODT powers and corresponding trap lifetimes are calculated. From these results we identify our best candidate for a magic wavelength trap. A lot of work, both theoretical and experimental, has been done for the dc polarizability of the $2 \ ^3S$ and $2 \ ^1S$ states (see Table \ref{table:dcpolarizabilities} for an overview). Therefore these are used as a benchmark for our calculations by also calculating the polarizabilities in the dc limit $(\lambda \to \infty)$, as discussed in Sec. \ref{sec:results}. Calculations of the ac polarizability of the $2 \ ^3S$ and $2 \ ^1S$ states \cite{Mitroy2,Chen1} states allows for comparison of the polarizability calculations at finite wavelengths.

Finally we present a simple extension to $^3$He which has a hyperfine structure that needs to be taken into account. Although different theoretical challenges arise due to the hyperfine interaction, we can get an estimation of the $^3$He magic wavelength candidates and show that they are equal to the $^4$He results approximately shifted by the hyperfine and isotope shift. 

\section{THEORY FOR $^4$He} \label{sec:theory}
For an atomic state with angular momentum \textit{J} and magnetic projection $M_J$, the polarizability $\alpha$ induced by an electromagnetic wave with polarization state $q$  ($q = 0, \pm 1$) and angular frequency $\omega$ due to a single opposite parity state is \cite{Sobel1}
\begin{widetext}
\begin{align}
\alpha^{(n)}(J,M_J,J',M_{J}',q) = 6 \pi \epsilon_0 c^3 (2J'+1) 
\begin{pmatrix}
  J & 1 & J' \\
  -M_J & q & M_J'
\end{pmatrix}
^2 \frac{A_{nJJ'}}{\omega_{nJJ'}^2 (\omega_{nJJ'}^2 - \omega^2)}. \label{eqn:indv_polarizability}
\end{align}
\end{widetext}
Here $\omega_{nJJ'}$ is the $2 \ ^{1,3}S_J \to n \ ^{1,3}P_{J'}$ transition frequency and $A_{nJJ'}$ the Einstein A coefficient of the transition. The term between two brackets represents the $3j$ symbol of the transition. The total polarizability $\alpha (J,M_J,q)$ is given by a sum over all opposite-parity states as
\begin{equation}
\alpha (J,M_J,q) = \sum_n \sum_{J'} \alpha^{(n)}(J,M_J,J',M_{J}',q). \label{eqn:polarizability}
\end{equation}
In a general way the polarizability $\alpha$ can be written as the sum of a scalar polarizability, independent of $M_J$, and a tensorial part describing the splitting of the $M_J$ levels \cite{Mitroy1,Mitroy3}. Within the $LS$ coupling scheme the tensor polarizability of the $2 \ ^3S_1$ and $2 \ ^1S_0$ states in $^4$He is zero and the polarizability is defined by averaging over all $M_J$ states and therefore independent of $M_J$. As our experimental work specifically concerns the spin-stretched $2 \ ^3S_1 \ (M_J = +1)$ state \cite{Rooij1,Notermans1}, Eqns. \ref{eqn:indv_polarizability} and \ref{eqn:polarizability} are used to calculate the polarizability for the $M_J = +1$ state assuming linearly polarized light ($q=0$). For $^3$He the calculations specifically concern the spin-stretched $2 \ ^3S \ F = 3/2 \ (M_F = +3/2)$ and $2 \ ^1S \ F = 1/2 \ (M_F = +1/2)$ states. 

The higher-order contribution to the Stark shift, the hyperpolarizability, is estimated using calculations of a similar system \cite{Takamoto1}. The contribution is many orders of magnitude smaller than the accuracy of our calculations and therefore neglected.

The summation in Eqn. \ref{eqn:polarizability} can be explicitly calculated for $2 \ ^{1,3}S \to n \ ^{1,3}P$ transitions up to $n = 10$, as accurate \textit{ab initio} energy level data and Einstein A coefficients are available \cite{Drake1}. Extrapolation of both the energy levels and the Einstein A coefficients is required to calculate contributions of dipole transition matrix elements with states beyond $n = 10$. A straightforward quantum defect extrapolation can be used to determine the energies using the effective quantum number $n^*$ \cite{Drake2}:
\begin{align}
n^* = n - \sum_{r = 0}^{\infty} \frac{\delta_r}{n^{*r}}, \label{eqn:effquantnumb}
\end{align}
where $\delta_r$ are fit parameters and the quantity $n - n^*$ is commonly referred to as the quantum defect. For both the singlet and triplet series, Eqn. \ref{eqn:effquantnumb} is used to fit the literature data up to $n = 10$ and to extrapolate to arbitrary $n$. This method is tested using a dataset provided by Drake \cite{Drake2}.

Extrapolation of the Einstein A coefficients is more complicated as there is no relation such as Eqn. \ref{eqn:effquantnumb} for Einstein A coefficients. Furthermore, the sum-over-states method does not provide straightforward extrapolation beyond the ionization limit, as the energy levels converge to the ionization limit for $n \to \infty$. Both problems can be solved by calculating the polarizability contribution of a single transition $2 \ ^3S_1 \to n \ ^3P_{J'}$ (or $2 \ ^1S_0 \to n \ ^1P_1$) as given in Eqn. \ref{eqn:polarizability} and defining the polarizability density per upper state energy interval as
\begin{align}
\frac{\Delta \alpha^{(n)}}{\Delta E} = \frac{2 \alpha^{(n)}}{E_{n+1}-E_{n-1}}, \label{eqn:poldensity}
\end{align}
which is evaluated at $E_n$. $E_{n+1}$ and $E_{n-1}$ are the energies of the neighbouring upper states with the same value of $J'$. The energies are given by the Rydberg formula $E_n(n^*) = E_{\text{IP}}- R_{\infty}/n^{*2}$, where $E_{\text{IP}}$ is the ionization potential of the ground state. For ease of notation we have omitted all the dependent variables of $\alpha^{(n)}$ as defined in Eqn. \ref{eqn:indv_polarizability}. The polarizability density is a function of energy and can not only be used to calculate the polarizability contribution from dipole transition matrix elements to highly excited (Rydberg) states, but additionally allows extrapolation beyond the ionization potential. Using the Rydberg formula, the polarizability density becomes
\begin{align}
\frac{\Delta \alpha^{(n)}}{\Delta E} = \frac{\alpha^{(n)}}{R_{\infty}} \frac{(n^{*2}-1)^2}{2 n^*}, \label{eqn:poldensity2}
\end{align}
where we have made the approximation that $n - n^*$ is constant for increasing $n$. This approximation already works better than $1\%$ for $n = 2$. In the limit $n \gg 1$, the polarizability contribution per energy interval can be written as
\begin{widetext}
\begin{align}
\frac{d \alpha^{(n)}}{d E} = \frac{6 \pi \epsilon_0 c^3}{R_{\infty}} (2J' + 1)
\begin{pmatrix}
  J & 1 & J' \\
  -M_J & q & M_J'
\end{pmatrix}
^2 \frac{C_{nJJ'}(n^*)}{\omega^2_{nJJ'}(\omega^2_{nJJ'} - \omega^2)}, \label{eqn:poldensity3} 
\end{align}
\end{widetext}
where we define 
\begin{align}
C_{nJJ'}(n^*) \equiv \frac{A_{nJJ'}(n^{*2}-1)^2}{2 n^*}. \label{eqn:Aextrap}
\end{align}
As there is no exact analytical model for $A_{nJJ'}$ as function of energy, the method of extrapolation is based on a simple low-order polynomial fit of the $C_{nJJ'}(n^*)$ as function of $E(n^*)$ for the $n \leq 10$ levels. The result is a function $C_{nJJ'}(E)$ that is used to extrapolate $A_{nJJ'}$ to arbitrary upper states and calculate the corresponding polarizability contributions. This method can be used to calculate the finite polarizability contributions of all Rydberg states for $n \to \infty$. As the general behaviour of the Einstein A coefficients is proportional to $n^{*-3}$ for the Rydberg states, $C_{nJJ'}(E)$ will have a finite value at the ionization potential indicating that contributions from the continuum have to be taken into account as well. As the extrapolation is a function of energy, it is extended beyond the ionization potential to calculate additional continuum contributions to the polarizability. This omits all higher order effects such as resonances to doubly-excited states or two-photon excitations into the continuum, and it should be considered as an approximation of the continuum. 

For a large enough quantum number $n$, the discrete sum-over-states method smoothly continues as an integration-over-states method following Eqn. \ref{eqn:poldensity3}. The ionization potential serves as a natural choice as the energy at which the calculation would switch from the discrete sum to the integration method. But even for large enough $n$ there is a negligible numerical error in varying the exact cutoff energy $E_c$ at which we switch between these methods. The calculation of the total polarizability is therefore performed using the sum-over-states method to an arbitrary cutoff at $E_c = E_{\text{IP}} - R_{\infty}/n_{max}^{*2}$ and continued with an integration over the remaining states as 
\begin{align}
\alpha^{\text{cont}} (J,M_J) = \sum_{J'} \int_{E_c}^{\infty} \frac{d \alpha^{(n)}}{d E} \text{d}E, \label{eqn:continuum}
\end{align}
where $E$ is the energy of the corresponding state. A low-order polynomial fit of Eqn. \ref{eqn:Aextrap} is used to calculate $d \alpha^{(n)} / d E$ such that the integral of Eqn. \ref{eqn:continuum} provides an analytical solution. The total polarizability is therefore easily calculated as a sum-over-states part and an analytical expression 
\begin{align}
\alpha (J,M_J)  = \alpha^{\text{cont}} + \sum_{n=1}^{n=n_{max}}  \sum_{J'} \alpha^{(n)}. \label{eqn:totpol} 
\end{align}

\section{Numerical uncertainties} \label{sec:numuncertainties}
In this section we discuss the sources of any numerical errors in our calculations, which are purely based on the technical execution of our method. The accuracy of our calculations due to our estimation of the continuum contribution will be discussed in Sec. \ref{sec:results} where our results are compared to other calculations.

The numerical convergence of Eqn. \ref{eqn:totpol} is tested by varying $n_{max}$. The polarizability converges as $n_{max}^{-2}$ and even for $n_{max} = 20$ the polarizability is within a fraction $10^{-4}$ of the polarizability calculated using $n_{max} = 5000$. The computation of Eqn. \ref{eqn:totpol} therefore poses no numerical problems.

A more crucial matter is the fact that our calculations are based on two extrapolations: that of the level energies and the Einstein A coefficients. For the  $n \leq 10$ levels in helium the \textit{ab initio} calculations of the level energies and Einstein A coefficients are used \cite{Drake1}. The higher level energies are extrapolated using Eqn. \ref{eqn:effquantnumb} and include up to fifth order ($r = 5$) contributions. Variation of the total number of orders ($r = 4,6$) or using a different dataset (such as the NIST database \cite{NIST1} as used in other recent work \cite{Mitroy2}) affects the polarizabilities at the $10^{-8}$ level and is negligible.

The limiting factor in the accuracy of the calculations is the choice of extrapolation of the Einstein A coefficients through extrapolation of $C_{nJJ'}(E)$. As mentioned before, no advanced methods are used to calculate transtion matrix elements to higher states or doubly excited states in the continuum. The heuristic approach we use instead, is to choose an extrapolation function that is smooth, continuous and provides a convergent integral in Eqn. \ref{eqn:continuum}. A number of different functions have been tried which provide a similar quality of the fit, and their effect on the calculation of the continuum contribution can lead to a polarizability shift which is a significant fraction of the continuum contribution itself. In our calculations this is the limiting factor in the accuracy of the calculated magic wavelengths. A second order polynomial function is chosen to extrapolate $C_{nJJ'}(E)$ as it has the additional advantage of providing an analytical solution of the continuum contributions.

The absolute accuracy of the calculations will be discussed in Sec. \ref{sec:uncertainties} and determines the accuracy given in the calculated magic wavelengths in Sec. \ref{sec:magicwavelengths}.

\section{RESULTS} \label{sec:results}
In order to discuss the absolute accuracy of the calculations, we first present our polarizabilities calculated in the dc limit $(\lambda \to \infty)$ as a lot of literature is available for these calculations. After comparison with the dc polarizabilities in Sec. \ref{sec:uncertainties}, the ac polarizabilities are given in Sec. \ref{sec:magicwavelengths} including the magic wavelengths at which they are equal for the $2 \ ^3S_1 \ (M_J = +1)$ and $2 \ ^1S_0$ states. Experimental characteristics, such as the required trapping power and scattering lifetime at the magic wavelengths, are estimated in order to discuss which magic wavelength candidate is most suitable for our experiment. In Sec. \ref{sec:tuneout} the tune-out wavelength (where the polarizability is zero) of the $2 \ ^3S_1$ state near 414~nm is compared to the result calculated by Mitroy and Tang \cite{Mitroy2}.

\subsection{dc polarizabilities} \label{sec:uncertainties}
An overview of previously calculated and measured dc polarizabilities for the $2 \ ^1S_0$ and $2 \ ^3S_1$ states of $^4$He is given in Table \ref{table:dcpolarizabilities} together with our results. For convenience the polarizabilities are given in atomic units $a_0^3$ ($a_0$ is the Bohr radius), but they can be converted to SI units through multiplication by $4 \pi \epsilon_0 a_0^3 \approx 1.64877 \times 10^{-41}$~JV$^{-2}$m$^{2}$. Furthermore, the dc polarizabilities are calculated using the common convention of averaging over all $M_J$ states and all possible polarizations $q$ \cite{Mitroy1}. 

There is general agreement between our results and previously calculated dc polarizabilities, but comparison with the work of Yan and Babb \cite{Yan1}, which provides the most accurate calculated dc polarizabilities to date, shows that both our $2 \ ^1S_0$ and $2 \ ^3S_1$ dc polarizabilities are slightly larger ($0.1\%$ and $0.6\%$, respectively). The difference is comparable to the uncertainty in the calculated continuum contributions as discussed in Sec. \ref{sec:numuncertainties}, and we conclude that our absolute accuracy is indeed limited by the exact calculation of the continuum contributions. It should be noted that the continuum contributions in the dc limit are $7.1 \ a_0^3$ and $3.6 \ a_0^3$, respectively. This only contributes $1\%$ to the total polarizability in contrast to e.g. ground-state hydrogen for which the continuum contribution is 20\% of the total polarizability \cite{Castillejo1}. 

\begin{table*}[tbp]
\caption{\label{table:dcpolarizabilities} Comparison of calculations and measurements of $M_J$-averaged dc polarizabilities of the $2 \ ^1S_0$ and $2 \ ^3S_1$ states in units of $a_0^3$.}
\begin{ruledtabular}
\begin{tabular}{lccc}
{Author (year)} & {Ref.} & {$2 \ ^1S_0$} & {$2 \ ^3S_1$}
\\ \hline
{Crosby and Zorn (1977) [Experiment]} & {\cite{Crosby1}} & {$729(88)$} & {$301(20)$}
\\
{Ekstrom \textit{et al.} (1995) [Experiment]} & {\cite{Ekstrom1,Molof1}} & {} & {$322(6.8)$}
\\ 
{Chung and Hurst (1966)} & {\cite{Chung1}} & {$801.95$} & {$315.63$}
\\
{Drake (1972)} & {\cite{Drake3}} & {$800.2$} & {$315.608$}
\\
{Chung (1977)} & {\cite{Chung2}} & {$801.10$} & {$315.63$}
\\
{Glover and Weinhold (1977)} & {\cite{Glover1}} & {$803.31$} & {$316.24$}
\\
{Lamm and Szabo (1980)} & {\cite{Lamm1}} & {$790.8$} & {$318.7$}
\\
{Bishop and Pipin (1993)} & {\cite{Bishop1}} & {} & {$315.631$}
\\
{R\'erat \textit{et al.} (1993)} & {\cite{Rerat1}} & {$803.25$} & {}
\\
{Chen (1995)} & {\cite{Chen1}} & {$800.31$} & {}
\\
{Chen and Chung (1996), \textit{B} Spline} & {\cite{Chen2}} & {} & {$315.630$}
\\
{Chen and Chung (1996), Slater} & {\cite{Chen2}} & {} & {$315.611$}
\\
{Yan and Babb (1998)} & {\cite{Yan1}} & {$800.316\,66$} & {$315.631\,468$}
\\
{Mitroy and Tang (2013), hybrid} & {\cite{Mitroy2}} & {} & {$315.462$}
\\
{Mitroy and Tang (2013), CPM} & {\cite{Mitroy2}} & {} & {$316.020$}
\\
{This work} & {} & {$801.19$} & {$317.64$}
\end{tabular}
\end{ruledtabular}
\end{table*}

\begin{table*}[tbp]
\caption{\label{table:magicwavelengths} Calculated magic wavelengths $\lambda_m$ for the $2~^3S_1~(M_J = +1)~\to~2~^1S_0$ transition with the corresponding differential polarizability slope $d\alpha/d\lambda$ and the absolute polarizability $\alpha$ at the magic wavelength. The last row gives the wavelength and polarizability at which we currently use our ODT. Additional columns give the laser beam power required to create a $5 \ \mu$K deep trap in the exact same crossed-beam geometry as currently employed and the corresponding lifetime of the gas in this geometry due to scattering from a nearby $2~^3S_1~\to~n~^3P_{0,1,2}$ transition. See the Appendix for details on those calculations.}
\begin{ruledtabular}
\begin{tabular}{c c d d c c}
\multicolumn{1}{c}{$\lambda_m \ \text{[nm]}$} & $d\alpha/d\lambda \ [a_0^3\text{/nm}]$ & \multicolumn{1}{c}{$\alpha \ [a_0^3]$} & \multicolumn{1}{c}{Laser power [W]} & \multicolumn{1}{c}{Lifetime [s]}& \multicolumn{1}{c}{Nearest transition}
\\ \hline
318.611 & $-7.00 \times 10^4$ & -809.2 & & &
\\
319.815 & {$-4.40 \times 10^3$} & 189.3 & 0.7 & 3 & {$2 \ ^3S_1 \to 4 \ ^3P_{0,1,2}$}
\\
321.409 & {$-5.38 \times 10^2$} & 55.3 & 2.3 & 6 & {$2 \ ^3S_1 \to 4 \ ^3P_{0,1,2}$}
\\
323.587 & {$-1.48 \times 10^2$} & 17.2 & 7.3 & 6 & {$2 \ ^3S_1 \to 4 \ ^3P_{0,1,2}$}
\\
326.672 & {$-5.48 \times 10^1$} & -1.2 & & &
\\
331.268 & {$-2.37 \times 10^1$} & -13.5 & & &
\\
338.644 & {$-1.08 \times 10^1$} & -24.2 & & &
\\
352.242 & {$-5.33$} & -39.0 & & &
\\
411.863 & {$-2.00$} & 4.5 & 28.0 & 4 & {$2 \ ^3S_1 \to 3 \ ^3P_{0,1,2}$}
\\ \hline
1557.3 & {$0.0$} & 603.8 & 0.2 & 205 & {$2 \ ^3S_1 \to 2 \ ^3P_{0,1,2}$}
\end{tabular}
\end{ruledtabular}
\end{table*}

\subsection{Magic wavelengths} \label{sec:magicwavelengths}
We have calculated the ac polarizabilities of the $2 \ ^1S_0$ and $2 \ ^3S_1 \ (M_J = +1)$ states in the range of 318~nm to 2.5~$\mu$m and an overview of the identified magic wavelengths is shown in Table \ref{table:magicwavelengths}. The slope of the differential polarizability is also given in order to estimate the sensitivity of the determined magic wavelength due to the accuracy of the calculated polarizabilities. Table \ref{table:magicwavelengths} furthermore provides the trapping beam power required to produce a trap depth of $5 \ \mu$K and the corresponding scattering lifetime (see the Appendix) to indicate the experimental feasibility of each magic wavelength.

\begin{figure*}[tbp]
	\begin{center}
		\includegraphics[width=0.82\textwidth]{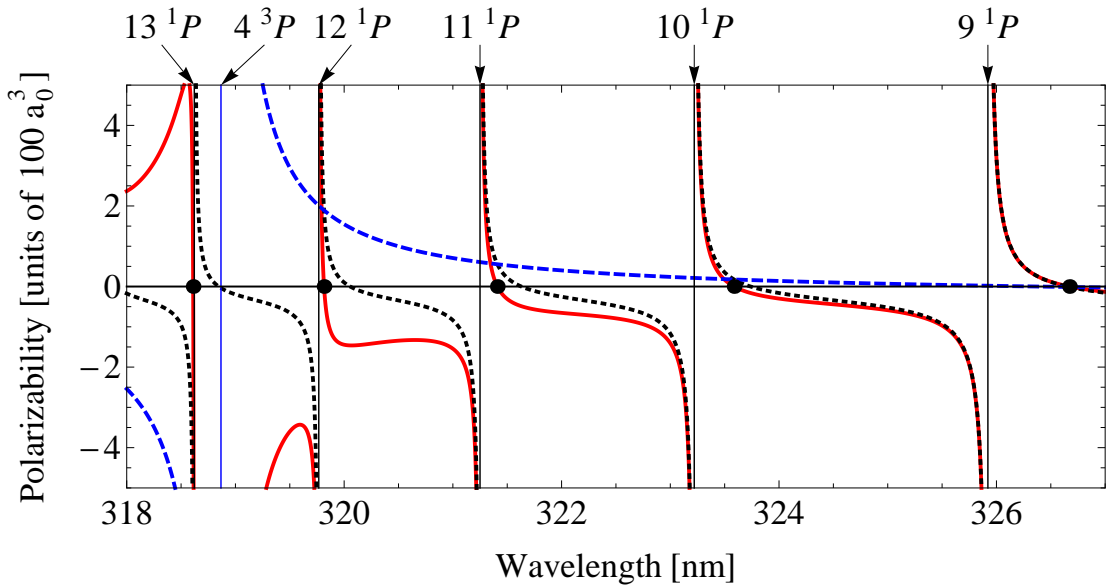}
		\caption{(Color online) Calculated polarizabilities of the $2 \ ^3S_1$ (dashed, blue) and $2 \ ^1S_0$ (dotted, black) states shown together with the differential polarizability (full, red) in the wavelength range 318-327~nm. The blue and black vertical lines indicate the positions of the $2 \ ^3S_1 \to 4 \ ^3P$ and the $2 \ ^1S_0 \to n \ ^1P \ (n = 9 - 13)$ transitions, respectively. There are five magic wavelengths (black dots) in this range, all listed in Table \ref{table:magicwavelengths}.}
		\label{fig:318-327nm}
	\end{center}
\end{figure*}

The magic wavelengths in the range 318-327~nm, as shown in Fig. \ref{fig:318-327nm}, are mainly due to the many resonances in the singlet series. The most promising magic wavelength for application in the experiment is at 319.815~nm, as the polarizability is large enough to provide sufficient trap depth at reasonable laser powers while the estimated scattering lifetime is still acceptable (see Table \ref{table:magicwavelengths}).

\begin{figure*}[tbp]
	\begin{center}
		\includegraphics[width=0.82\textwidth]{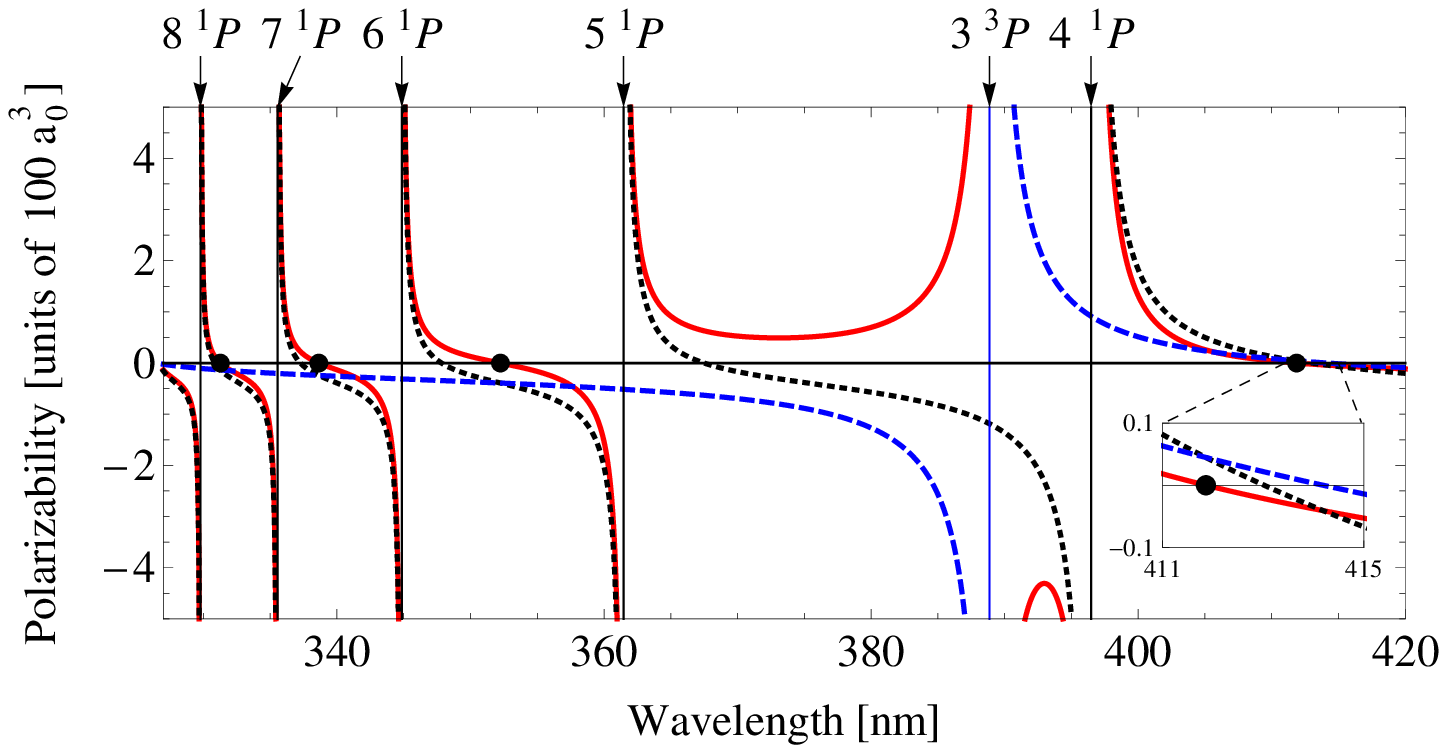}
		\caption{(Color online) Calculated polarizabilities of the $2 \ ^3S_1$ (dashed, blue) and $2 \ ^1S_0$ (dotted, black) states shown together with the differential polarizability (full, red) for wavelengths ranging from 327 nm to 420 nm. The blue and black vertical lines indicate the positions of the $2 \ ^3S_1 \to 3 \ ^3P$ and the $2 \ ^1S_0 \to n \ ^1P \ (n = 4 - 8)$ transitions, respectively. There are four magic wavelengths (black dots) in this range, all listed in Table \ref{table:magicwavelengths}. The inset shows the wavelength region 411-415 nm, displaying the magic wavelength at 411.863~nm and the tune-out wavelength of the $2 \ ^3S_1$ state at 414.197~nm.}
		\label{fig:327-420nm}
	\end{center}
\end{figure*}

\begin{figure*}[tbp]
	\begin{center}
		\includegraphics[width=0.82\textwidth]{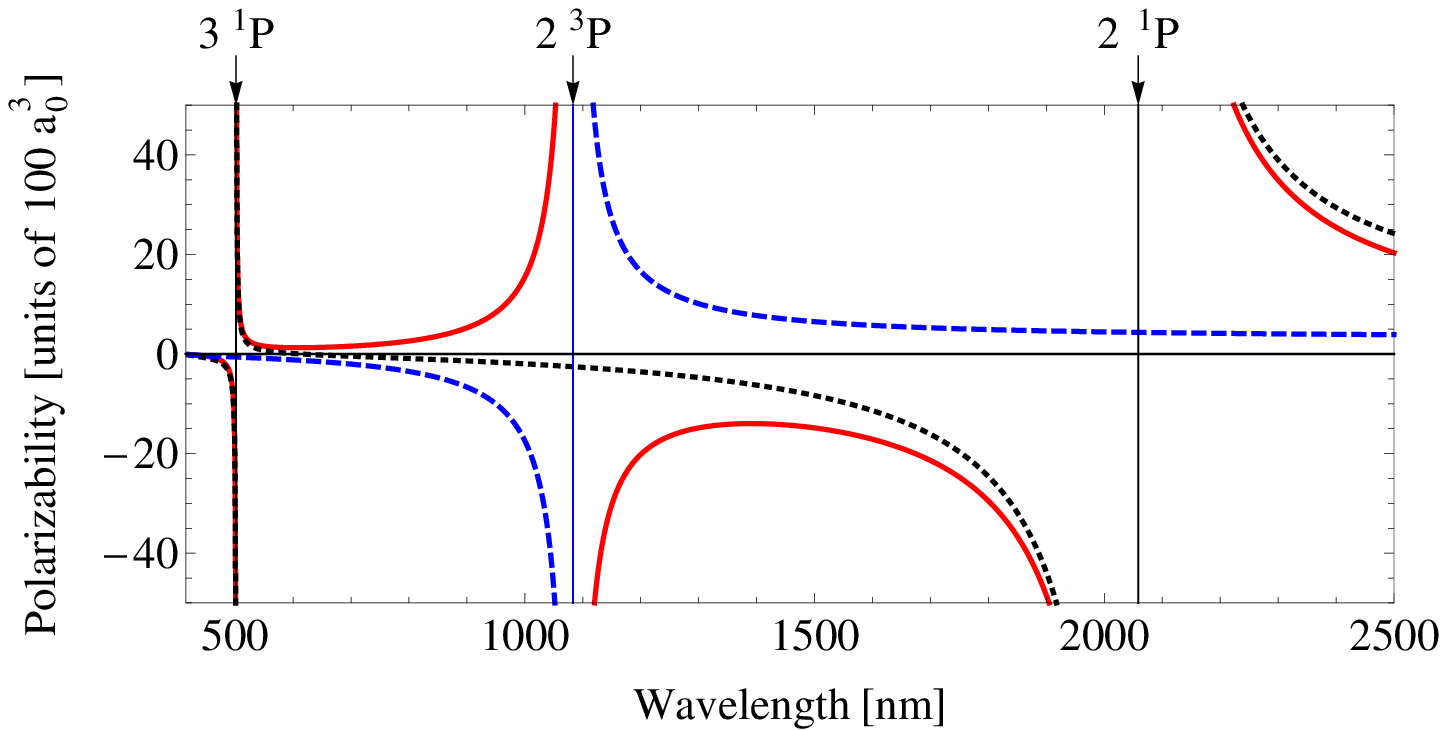}
		\caption{(Color online) Calculated polarizabilities of the $2 \ ^3S_1$ (dashed, blue) and $2 \ ^1S_0$ (dotted, black) states shown together with the differential polarizability (full, red) for wavelengths ranging from 420 nm to $2.5 \ \mu$m. The blue and black vertical lines indicate the positions of the $2 \ ^3S_1 \to 2 \ ^3P$ and the $2 \ ^1S_0 \to n \ ^1P \ (n = 2,3)$ transitions, respectively. There are no magic wavelengths in this range and the polarizabilities converge to the dc polarizabilities for $\lambda > 2.5 \ \mu$m.}
		\label{fig:420-2500nm}
	\end{center}
	\end{figure*}

The magic wavelengths at 318.611~nm and 326.672~nm are not useful for our experiment as the absolute $2 \ ^3S_1$ polarizability is negative and therefore a focused laser beam does not provide a trapping potential. There are more magic wavelengths for $\lambda < 318.611$~nm, but the polarizability of the $2 \ ^3S_1$ state will stay negative until the ionization wavelength of the $2 \ ^1S$ state around 312~nm. In the range 327-420~nm, shown in Fig. \ref{fig:327-420nm}, there are four more magic wavelengths. The magic wavelength at 411.863~nm, previously predicted with nm accuracy \cite{Eyler1}, is the only one in this region with a small yet positive $2 \ ^3S_1$ polarizability (see inset in Fig. \ref{fig:327-420nm}). There are no more magic wavelengths in the range 420~nm-2.5~$\mu$m, which is shown in Fig. \ref{fig:420-2500nm}, and the polarizabilities converge to the dc polarizabilities for $\lambda > 2.5 \ \mu$m.

The ac polarizability of the $2 \ ^1S_0$ state can be compared to previous polarizability calculations from dc to 506~nm \cite{Chen1}. Combined with the dc polarizability comparison and the tune-out wavelength result for the $2 \ ^3S_1$ state, as discussed in the Sect. \ref{sec:tuneout}, we find that the accuracy of our calculations is limited by the exact calculation of the continuum contributions. We note that around 320~nm the absolute continuum contributions ($26 \ a_0^3$ and $5.5 \ a_0^3$ for the $2 \ ^1S$ and $2 \ ^3S$ states, respectively) and the corresponding uncertainty have increased, as the shorter wavelengths are closer to the $2 \ ^1S$ ionization limit at 312~nm. The uncertainty in the absolute value of the polarizabilities translates to an uncertainty in the absolute value of the magic wavelength through the slope $d\alpha / d\lambda$ of the differential polarizability at the zero crossing. For the magic wavelength at 319.815~nm this gives a frequency uncertainty of 10~GHz (0.003~nm), yet for the magic wavelength near 412~nm the uncertainty is approximately 1~nm due to the very small slope at the zero crossing. However, the latter magic wavelength is not suitable for our experiment as the absolute polarizability is very small.


\subsection{Tune-out wavelength of the $2 \ ^3S_1$ state} \label{sec:tuneout}
The zero crossings of the absolute polarizability of a single state occur at so-called tune-out wavelengths. Mitroy and Tang calculated several tune-out wavelengths for the $2 \ ^3S_1$ state \cite{Mitroy2}, of which the candidate at 413.02~nm is the most sensitive to the absolute value of the polarizability due to a very small slope at the zero crossing. We find this tune-out wavelength at 414.197~nm (see inset in Fig. \ref{fig:327-420nm}), which is considerably larger. However, the slope of the polarizability at the zero crossing can be used to calculate that the difference in tune-out wavelength is equivalent to a difference in the calculated absolute polarizabilities. Comparison of the calculated dc polarizabilities (see Table \ref{table:dcpolarizabilities}) shows a similar difference, so within a constant offset of the absolute polarizability our tune-out wavelength is in agreement with Mitroy and Tang's result. 

\section{EXTENSION TO $^3$He}
The $2~^3S \to 2~^1S$ transition is also measured in $^3$He in order to determine the isotope shift of the transition frequency \cite{Rooij1}. Hence a magic wavelength trap for $^3$He will be required as well. As $^3$He has a nuclear spin ($I = 1/2$), the measured hyperfine transition is $2~^3S~F = 3/2~(M_F = +3/2) \to 2~^1S~F = 1/2~(M_F = +1/2)$ and the magic wavelengths need to be calculated for these two spin-stretched states.

The mass-dependent (isotope) shift of the energy levels is taken into account by using $^3$He energy level data \cite{Morton2} and recalculating the quantum defects using Eqn. \ref{eqn:effquantnumb}. The Einstein A coefficients of the transitions also change due to the different reduced mass of the system \cite{Drake1}, but this effect is negligible compared to the accuracy of the calculations. In total, the mass-dependent shift of the magic wavelengths is dominated by the shift of the nearest transitions and is approximately -45~GHz.

The fine-structure splitting decreases as $1/n^3$ whereas the hyperfine splitting converges to a constant value for increasing $n$ \cite{Vassen1}. In this regime the $(LS)JIF$ coupling scheme is not the best coupling scheme because $J$ is no longer a good quantum number. Instead an alternative coupling scheme is used which first couples the nuclear spin quantum number $I$ and total electron spin $S$ to a new quantum number $K$ \cite{Sulai1}. This new quantum number $K$ then couples to $L$ to form the total angular momentum $F$. In this coupling scheme the transition strengths can be calculated with better precision compared to the $(LS)JIF$ coupling scheme, and can be applied for states with $n \geq 3$. Although this coupling scheme does not work perfectly for $n = 2$ (which in any case is far-detuned from the magic wavelengths), it provides an estimate of the transition strengths that is sufficiently accurate for our purposes. 

For increasing $n$, the strong nuclear spin interaction with the $1s$ electron becomes comparable with the exchange interaction between the $1s$ and $np$ electrons \cite{Vassen1}. This leads to mixing of the singlet and triplet states as the total electron spin $S$ is no longer a good quantum number. The solution requires exact diagonalization of the Rydberg states, which provides the singlet-triplet mixing and the energy shifts of the states. The mixing parameter is then used to correct the Einstein A coefficients and the energies of the states. Although this is implemented in the calculations, these corrections lead to shifts in the magic wavelengths that are below the absolute accuracy of the calculations. 

Due to the two hyperfine states of $^3$He$^+$ in the $1s$ ground state, there are two Rydberg series in the $^3$He atom. For even higher $n$ than discussed before, this leads to mixing of Rydberg states with different $n$ \cite{Vassen1}. The resulting shifts in the polarizabilities are well below the accuracy of the calculations and are therefore neglected. 

Using the aforementioned adaptations, the polarizability of the $2~^3S~F = 3/2~(M_F = +3/2)$ and $2~^1S~F=1/2~(M_F = +1/2)$ states can be calculated using Eqn. \ref{eqn:indv_polarizability}, but with substituted quantum numbers $(J,M_J \to F,M_F)$, Einstein A coefficients and transition frequencies. The numerical calculation of the polarizabilities and discussion of the numerical accuracies is similar to the $^4$He case. An additional uncertainty of $1.0 \ a_0^3$ is added in the calculation of the polarizabilities of the $^3$He states based on a conservative estimate of the shifts caused by the hyperfine interaction. It should be noted that the states of interest, $2~^1S$ and $2~^3S$, both have angular momentum $L = 0$ and both are in the fully spin-stretched state. Therefore neither $^3$He nor $^4$He has a tensor polarizability for the states discussed in this paper.

A comparison between the $^4$He and $^3$He magic wavelengths is presented in Table~\ref{tab:He3magic}. Magic wavelengths up to 330~nm are all shifted by the isotope shift with small corrections due the abovementioned effects. The frequency difference between the two isotopes (third column of Table \ref{tab:He3magic}) grows with increasing wavelengths because $d\alpha / d\lambda$ decreases and the results become more sensitive to the absolute accuracy ($1.0 \ a_0^3$) of the calculations, as can be seen from the growing uncertainties associated with the shifts. The isotope shifts for magic wavelengths with $\lambda > 324$~nm have been omitted in Table \ref{tab:He3magic} as they are not useful due to the large relative uncertainty.

The difference of the magic wavelengths between the two isotopes is well within the tuning range of our designed laser system near 320~nm. Furthermore there is no significant change in the absolute polarizability or the slope $d\alpha / d\lambda$ at the magic wavelengths. This means that an ODT at these wavelengths has a comparable performance for either isotope.

\begin{table}
\caption{Comparison of magic wavelengths $\lambda_m$ calculated for the $^4$He $2 \ ^3S_1 \ (M_J = +1) \to 2 \ ^1S_0$ and $^3$He $2 \ ^3S \ F = 3/2 \ (M_F = +3/2) \to 2 \ ^1S \ F = 1/2 \ (M_F = +1/2)$ transitions and the corresponding frequency shift. The uncertainty in the shift is due to the additional $1.0 \ a_0^3$ absolute uncertainty in the polarizabilities of $^3$He.}
\begin{ruledtabular}
\begin{tabular}{c c c}
\multicolumn{2}{c}{$\lambda_m \ \text{[nm]}$} & \multicolumn{1}{c}{Shift [GHz]} 
\\
\multicolumn{1}{c}{$^4$He} & \multicolumn{1}{c}{$^3$He} & 
\\ \hline
318.611 & 318.626 & $-45.03(4)$
\\
319.815 & 319.830 & $-43.1(7)$
\\ 
321.409 & 321.423 & $-38(5)$
\\
323.587 & 323.602 & $-4(2) \times 10^1$
\end{tabular}
\end{ruledtabular}
\label{tab:He3magic}
\end{table}

\section{CONCLUSION}
We have calculated the dc and ac polarizabilities of the $2 \ ^1S$ and $2 \ ^3S$ states for both $^4$He and $^3$He in the wavelength range of 318~nm to $2.5 \ \mu$m and determined the magic wavelengths at which these polarizabilities are equal for either isotope. The accuracy of our simple method is limited by the extrapolation of the polarizability contributions of the continuum states. This is less than achievable through more sophisticated methods which calculate the transition matrix elements explicitly. However, the purpose of this paper is to show that using a simple extrapolation method it is possible to achieve an accuracy on the order of 10~GHz for the magic wavelengths that are of experimental interest, which is required to design an appropriate laser system for the required wavelengths.

Most experimentally feasible magic wavelength candidates are in the range of 319-324 nm, as the absolute polarizability of the $2 \ ^3S_1$ state in this range is positive and large enough to create reasonable ($\sim\mu$K) trap depths in a crossed-beam ODT with a few Watts of laser power. The estimated scattering rates at these wavelengths and intensities are low enough to perform spectroscopy on the doubly-forbidden $2 \ ^3S \to 2 \ ^1S$ transition.

The calculations are extended to also calculate magic wavelengths in $^3$He. Although the hyperfine structure, which is absent in $^4$He, leads to complications in the calculation of the polarizabilities, these effects are very limited for the $2 \ ^1S$ and $2 \ ^3S$ states. The magic wavelengths of interest, around 320~nm, are shifted relative to the $^4$He magic wavelengths by predominantly the isotope shift. 

\begin{acknowledgements}
This work is part of the research programme of the Foundation for Fundamental Research on Matter (FOM), which is financially supported by the Netherlands Organisation for Scientific Research (NWO).
\end{acknowledgements}

\appendix* \section{a crossed-beam optical dipole trap}

An overview of optical dipole traps (ODTs) and the equations used in this Appendix can be found in \cite{Grimm1}. The depth $U$ of a crossed-beam ODT, as currently used in our experiment \cite{Rooij1,Notermans1}, is
\begin{align}
U = 2 \frac{\alpha}{2 \epsilon_0 c} \frac{2 P}{\pi w_0^2}, \label{eqn:trapdepth}
\end{align}
where $\alpha$ is the polarizability of the $2 \ ^3S_1 \ (M_J = +1)$ state, $P$ the power of the incident trapping laser beam and $w_0$ the beam waist. In our experiment, the first ODT beam is reused by refocusing it through the original focus ($w_0 \approx 85 \ \mu$m) at an angle of $19^{\circ}$ with respect to the original beam. At the currently used ODT wavelength of 1557~nm the polarizability is $\alpha = 603.8 \ a_0^3$ (see Table \ref{table:magicwavelengths}) which gives a trap depth of approximately $5 \ \mu$K at an ODT beam power of $P = 210$~mW. In Table \ref{table:magicwavelengths} we used Eqn. \ref{eqn:trapdepth} to calculate the trapping power at the different magic wavelengths corresponding to a trap depth of $5 \ \mu$K to indicate the required beam power that should be produced at that magic wavelength.

As a good approximation of the lifetime of the atoms in the ODT due to scattering, one can take the nearest transition into account to calculate the corresponding scattering rate. The scattering rate $\Gamma_{sc}$ is 
\begin{align}
\Gamma_{sc} = \frac{6 \pi c^2 \omega^3}{\hbar} \Bigg( \frac{\Gamma}{\omega_0^2(\omega_0^2-\omega^2)}  \Bigg)^2 I_0,
\end{align}
where $I_0$ is the total intensity of the light, $\omega$ the angular frequency of the trapping light and $\omega_0$ and $\Gamma$ the transition frequency and linewidth (all in $\textit{rad} \ s^{-1}$). The nearest transitions are given in Table \ref{table:magicwavelengths}, and the lifetime $1/\Gamma_{sc}$ is given for each magic wavelength trap using the required trapping beam power calculated to provide a $5 \ \mu$K deep trap.

\bibliography{HeliumMagicWavelength_references}
\bibliographystyle{apsrev4-1}

\end{document}